\documentclass[]{article}
\usepackage{multicol}
\usepackage{graphicx}
\usepackage{amssymb}
\usepackage{epsfig}
\usepackage[greek,english]{babel}
\usepackage[iso-8859-7]{inputenc}
\usepackage{feynmp}
\usepackage{caption}
\usepackage{subcaption}
\usepackage{amsmath}
\usepackage{float}
\usepackage{verbatim}
\usepackage{wrapfig}
\usepackage{textcomp}
\usepackage{color}
\usepackage{sidecap}
\usepackage{cancel}
\usepackage{amssymb}
\usepackage{kerkis}
\usepackage{youngtab}
\usepackage{stackrel}
\usepackage[toc,page]{appendix}
\usepackage{multirow}
\usepackage{bm}
\usepackage{cite}
\usepackage[absolute]{textpos}
\usepackage[stable]{footmisc}
\newcommand{\newc}{\newcommand}
\newc{\ra}{\rightarrow}
\newc{\lra}{\leftrightarrow}
\newc{\ov}{\overline}
\newc{\pa}{\partial}

\newc{\be}{\begin{equation}}
\newc{\ee}{\end{equation}}

\newc{\ba}{\begin{eqnarray}}
\newc{\ea}{\end{eqnarray}}

\newc{\n}{\nu}
\newc{\D}{\Delta}

\newc{\la}{\lambda}
\newc{\e}{\epsilon}
\newc{\nn}{\nonumber}

\def\bea{\begin{eqnarray}}
\def\eea{\end{eqnarray}}
\begin{document}
\begin{textblock}{5}(11,1)
\mbox{CERN-PH-TH-2014-232}
\end{textblock}
\title{Towards realistic models from Higher-Dimensional
theories with Fuzzy extra dimensions\thanks{Based on a talk
presented at the International Conference "Quantum Field Theory and
Gravity (QFTG'14)" (Tomsk, July 28 - August 3, 2014) by G.Z.
(invited main speaker).}}
\author{D.Gavriil\textsuperscript{1}, G.Manolakos\textsuperscript{1}, G.Zoupanos\textsuperscript{1,2}}\date{} \maketitle
\begin{center}
\itshape\textsuperscript{1}Physics Department, National Technical
University,\\ Zografou Campus, GR-15780 Athens, Greece\\
\vspace{0.15cm} \textsuperscript{2} Theory Group, Physics Department \\
Cern, Geneva, Switzerland
\end{center}
\begin{center}
\emph{E-mails:
d.gavriil@yahoo.gr\,,\,gmanol@central.ntua.gr\,,\,George.Zoupanos@cern.ch}
\end{center}
\begin{abstract}
We briefly review the Coset Space Dimensional Reduction (CSDR)
programme and the best model constructed so far and then we present
some details of the corresponding programme in the case that the
extra dimensions are considered to be fuzzy. In particular, we
present a four-dimensional $\mathcal{N} = 4$ Super Yang Mills
Theory, orbifolded by $\mathbb{Z}_3$, which mimics the behaviour of
a dimensionally reduced $\mathcal{N} = 1$, 10-dimensional gauge
theory over a set of fuzzy spheres at intermediate high scales and
leads to the trinification GUT $SU(3)^3$ at slightly lower, which in
turn can be spontaneously broken to the MSSM in low scales.
\end{abstract}
\begin{multicols}{2}

\section{Introduction}

$\,\,\,\,\,\,$ Since 1970's there has been an intense pursuit of
unification, that is the establishment of a single theoretical model
describing all interactions. Profound research activity has resulted
in two very interesting frameworks, namely Superstring Theories
\cite{gsw} and Non-Commutative Geometry \cite{connes}. Both
approaches, although developing independently, share common
unification targets and aim at exhibiting improved renormalization
properties in the ultraviolet regime as compared to ordinary field
theories. Moreover, these two (initially) different frameworks were
bridged together after realizing that a Non-Commutative gauge theory
can describe the effective physics on D-branes whilst a
non-vanishing background antisymmetric field is present
\cite{protianafora}.

Significant progress has recently been made regarding the
dimensional reduction of the $E_8\times E_8$ Heterotic String using
non-symmetric coset spaces \cite{LopesCardoso:2002hd}-
\cite{Klaput:2012vv}, in the presence of background fluxes and
gaugino condensates. It is widely known that the large number of
Standard Model's free parameters which enter the theory, because of
the ad hoc introduction of the Higgs and Yukawa sectors, is a major
problem demanding solution. This embarrassment can be overcome by
considering that those sectors originate from a higher dimensional
theory. Various frameworks starting with the Coset Space Dimensional
Reduction (CSDR) \cite{csdr10, csdr11, csdr12} and the
Scherk-Schwarz \cite{scherk-schwarz} reduction schemes suggest that
unification of the gauge and Higgs sectors can take place making use
of higher dimensions. This means that the four-dimensional gauge and
Higgs fields are the surviving components of the reduction procedure
of the gauge fields of a pure higher-dimensional gauge theory.
Furthermore, the addition of fermions in the higher-dimensional
gauge theory leads naturally (after CSDR) to Yukawa couplings in
four dimensions. The last step in this unified description in high
dimensions is to relate the gauge and fermion fields, which can be
achieved by demanding that the higher-dimensional gauge theory is
$\mathcal{N}=1$ supersymmetric, i.e. the gauge and fermion fields
are members of the same vector supermultiplet. In order to maintain
an $\mathcal{N}=1$ supersymmetry after dimensional reduction,
Calabi-Yau (CY) manifolds serve as suitable compact internal spaces
\cite{CY}. However, the moduli stabilization problem that arose, led
to the study of compactification with fluxes (for reviews see e.g.
\cite{grana}). Within the context of flux compactification, the
recent developments suggested the use of a wider class of internal
spaces, called manifolds with SU(3)-structure. The latter class of
manifolds admits a nowhere-vanishing, globally-defined spinor, which
is covariantly constant with respect to a connection with torsion
and not with respect to the Levi-Civita connection as in the CY
case. Here we focus on an interesting class of SU(3)-structure
manifolds called nearly-K{\"a}hler manifolds.

The homogeneous nearly-K{\"a}hler manifolds in six dimensions have
been classified in \cite{Butruille} and they are the three
non-symmetric coset spaces $G_2/SU(3)$, $Sp(4)/(SU(2)\times
U(1))_{non-max}$ and $SU(3)/U(1)\times U(1)$ and the group manifold
$SU(2)\times SU(2)$. The latter cannot lead to chiral fermions in
four dimensions and therefore, for our purposes, it is ruled out of
further interest. It is worth noting that four-dimensional theories
resulting from the dimensional reduction of ten-dimensional
$\mathcal{N}=1$ supersymmetric gauge theories over non-symmetric
coset spaces, contain terms which could be interpreted as soft
scalar masses. Here we will briefly describe the dimensional
reduction of the $\mathcal{N}=1$ supersymmetric $E_8$ gauge theory
over the nearly-K{\"a}hler manifold $SU(3)/U(1)\times U(1)$. More
specifically, an extension of the Minimal Supersymmetric Standard
Model (MSSM) was derived by dimensionally reducing the $E_8\times
E_8$ gauge sector of the heterotic string \cite{Irges-Zoupanos}.

Non-Commutative geometry is considered as an appropriate framework
for regularizing quantum field theories, or even better, building
finite ones. Unfortunately, constructing quantum field theories on
Non-Commutative spaces is much more difficult than expected and,
furthermore, they present problematic ultraviolet features
\cite{filk6}, see however \cite{groose7} and \cite{groose8}. In the
beginning, several models of the type of Standard Model were built
making use of the Seiberg-Witten map, but they could only be
considered as effective theories which were also lacking
renormalizability. A more promising use of Non-Commutative geometry
in particle physics occurred after the suggestion that it would
describe the extra dimensions \cite{aschieri9}; see also
\cite{schupp}. This proposal motivated the construction of
higher-dimensional models which present many interesting features
i.e. renormalizability, potential predictivity, etc.

In this framework has been developed a higher-dimensional gauge
theory in which those extra dimensions are described by fuzzy spaces
\cite{aschieri9}, i.e. matrix approximations on smooth manifolds.
The first step was to find a manifold on which one would construct a
higher-dimensional gauge theory. The appropriate one was the product
of Minkowski space and a fuzzy coset space $(S/R)_F$. Afterwards, in
order to achieve the necessary dimensional reduction, was made use
of the CSDR scheme, which is described in the previous section.
Although the reduction is performed using the CSDR programme, there
is a significant difference between the ordinary and the fuzzy
version: the four-dimensional gauge group that appears in the first,
between the geometrical and the spontaneous breaking due to the
four-dimensional Higgs fields, does not appear in the latter. In the
fuzzy CSDR scheme, the spontaneous symmetry breaking occurs after
solving the fuzzy CSDR constraints, resulting in a non-zero minimum
of the four-dimensional potential. Thus, in four dimensions, there
remains only one scalar field, the physical Higgs field, which does
survive the spontaneous symmetry breaking. In the same way,
regarding the Yukawa sector, we have the welcoming results of
massive fermions as well as interactions among the physical Higgs
field and fermions (Yukawa interactions). We conclude that in order
to be able to reproduce the spontaneous symmetry breaking of the SM
in this framework, one would have to consider large extra
dimensions. A determinant difference between ordinary and fuzzy CSDR
is that a non-Abelian gauge group $G$ is not necessary in the
higher-dimensions theory. The non-Abelian gauge theories in four
dimensions can originate from a $U(1)$ group in the
higher-dimensional theory.

These theories are equipped with a very strong advantage when
compared to the rest higher-dimensional ones, that is
renormalizability. Arguments leading to this result are given in
\cite{aschieri9}, but the strongest one was given after examining
the issue from a different perspective. In a detailed analysis, it
was established a renormalizable four-dimensional $SU(N)$ gauge
theory in which we assigned a scalar multiplet which dynamically
develops fuzzy extra dimensions, forming a fuzzy sphere
\cite{aschieri13}. The model develops non-trivial vacua which are
interpreted as 6-dimensional gauge theory, in which geometry and
gauge group depend on the parameters that are present in the initial
Langrangian. We result with a finite tower of massive Kaluza-Klein
modes, a result consistent with a dimensionally reduced
higher-dimensional gauge theory. This model presents many
interesting features. First, the extra dimensions are generated
dynamically by a geometrical mechanism. This feature is based on a
result from non-commutative gauge theory, namely that solutions of
matrix models can be interpreted as non-commutative, or fuzzy,
spaces. The above mechanism is very generic and does not need
fine-tuning, which means that supersymmetry is not involved. In the
renormalizable quantum field theory framework, this constitutes a
realization of the concepts of compactification and dimensional
reduction. Moreover, since it is a large $N$ gauge theory, every
analytical technique of this context should be available to be
applied. More specifically, it proves that the general gauge group
in low energies is $SU(n_1)\times SU(n_2)\times U(1)$ or $SU(n)$. In
this model, gauge groups that are formed by more than two simple
groups (apart from $U(1)$) are not observed.

The features that emerged from the above mechanism are quite
appealing, suggesting the construction of phenomenologically viable
models in particle physics. When addressed to this direction, one
encounters a severe problem, that is the chiral-fermion assignment
in four dimensions. The best candidates of the above category of
models, when it comes to inserting the fermions, are theories with
mirror fermions in bi-fundamental representations of the low energy
gauge group \cite{steinacker14}. Detailed studies on fermionic
sectors of models, which obey the mechanism of dynamical generation
of the extra dimensions with the fuzzy sphere or a product of two
fuzzy spheres, showed that when extrapolating to low-energy, the
fermionic sector of the theory consists of two mirror sectors, even
after the inclusion of the magnetic fluxes on the two fuzzy spheres
\cite{chatzistavrakidis15}. Although the presence of mirror fermions
does not exclude the possibility to obtain phenomenologically viable
models \cite{maalampi16}, it is certainly preferred to end up with
exactly chiral fermions. This is achieved by extending the above
context and by inserting an additional structure which is based on
orbifolds. Specifically, a $\mathbb{Z}_3$ orbifold projection of a
$\mathcal{N}=4$ $SU(3N)$ SYM theory leads to a $\mathcal{N}=1$
supersymmetric theory with the gauge group being the $SU(3)^3$
\cite{chatzistavrakidis17}. In order to obtain specific vacua in the
$\mathcal{N}=1$ theory, required by interpreting the theory as
resulting from fuzzy extra dimensions, one is normally obliged to
introduce soft breaking supersymmetric terms. This induces the
dynamical generation of twisted fuzzy spheres. The introduction of
such soft breaking terms seems necessary in order to build
phenomenologically viable supersymmetric theories, with MSSM being a
very leading case. The vacua that emerge give rise to models which
preserve the features described above, but also they accommodate a
chiral low-energy spectrum. The most appealing chiral models of this
kind are $SU(4)\times SU(2)\times SU(2)$, $SU(4)^3$ and $SU(3)^3$.
The most interesting of those unified theories seems to be the
latter, which is described by the trinification group. In addition,
this theory can be upgraded to a two-loop finite theory (for reviews
see \cite{inspire2}, \cite{inspire4}, \cite{heinemeyer18},
\cite{ma19}) and moreover it is able to make testable predictions
\cite{ma19}. Therefore, we conclude that fuzzy extra dimensions can
be used in constructing chiral, renormalizable and
phenomenologically viable field-theoretical models.

\section{The Coset Space Dimensional Reduction}

 In the Coset Space Dimensional Reduction (CSDR) scheme
(see \cite{csdr10, csdr11, csdr12} for a detailed exposition) one
starts with a Yang-Mills-Dirac Lagrangian , with gauge group G,
defined on a D-dimensional spacetime $M^D$, which is compactified to
$M_{4}\times S/R$ with $S/R$ a coset space. S acts as a symmetry
group on the extra coordinates and both S and its subgroup R, are
Lie groups. As far as the most general S-invariant metric is
concerned, it is always diagonal and depends on the number of radii
that each space admits. Regarding the coset of our interest
($SU(3)/U(1)\times U(1)$) three radii $R_1, R_2, R_3$ are
introduced. According to the CSDR framework, an S-transformation of
the extra d coordinates is a gauge transformation of the fields that
are defined on $M_{4}\times S/R$, thus a gauge invariant Lagrangian
written on this space is independent of the extra coordinates.
Fields defined in this way are called symmetric. The initial gauge
field $A_M(x,y)$ is split into its components $A_\mu(x,y)$ and
$A_a(x,y)$, corresponding to $M^4$ and $S/R$ respectively. Consider
the action of a D-dimensional Yang-Mills theory with gauge group G,
coupled to fermions defined on a manifold $M^D$ compactified on
$M_{4}\times S/R$, $D=4+d, d=dimS-dimR$:

 \vspace*{0.3cm}
\begin{align}
A&=\int d^4xd^dy\sqrt{-g}\nonumber\\
&\qquad\left[-\frac{1}{4}Tr(F_{MN}F_{K\Lambda})g^{MK}g^{N\Lambda}+\frac{i}{2}\bar
\Psi \Gamma^MD_{M}\Psi\right]\,,
\end{align}
\noindent where $D_{M}=\partial_{M}-\theta_{M}-A_{M}$, with
$\theta_{M}=\frac{1}{2}\theta_{MN\Lambda}\Sigma^{MN}$ the
spin-connection of $M^{D}$,
$F_{MN}=\partial_MA_N-\partial_NA_M-[A_M,A_N]$, where M, N run over
the D-dimensional space and $A_M$ and $\Psi$ are D-dimensional
symmetric fields. Let $\xi^{\alpha}_{A}, (A= 1,...,dimS$ and $\alpha
=dimR+1,...,dimS$ the curved index$)$ be the Killing vectors which
generate the symmetries of $S/R$ and $W_A$ the compensating gauge
transformation associated with  $\xi_A$. The requirement that
transformations of the fields under the action of $S/R$ are
compensated by gauge transformations, is expressed by the following
constraint equations for scalar $\phi$, vector $A_{\alpha}$ and
spinor $\psi$ fields on $S/R$, \vspace*{0.3cm}
\begin{gather}
\delta_A\phi=\xi^{\alpha}_{A}\partial_\alpha\phi=D(W_A)\phi, \\
\delta_AA_{\alpha}=\xi^{\beta}_{A}\partial_{\beta}A_{\alpha}+\partial_{\alpha}\xi^{\beta}_{A}A_{\beta}=\partial_{\alpha}W_A-[W_A,A_{\alpha}], \\
\delta_A\psi=\xi^{\alpha}_{A}\partial_\alpha\psi-\frac{1}{2}G_{Abc}\Sigma^{bc}\psi=D(W_A)\psi\,,
\end{gather}
where $W_A$ depend only on internal coordinates $y$ and $D(W_A)$
represents a gauge transformation in the appropriate representation
of the fields.

The constraints (2)-(4) provide us \cite{csdr10, csdr11} with the
four-dimensional unconstrained fields as well as with the gauge
invariance that remains in the theory after dimensional reduction.
The analysis of these constraints implies that the components
$A_\mu(x,y)$ of the initial gauge field $A_M(x,y)$ become, after
dimensional reduction, the four dimensional gauge fields and
furthermore they are independent of $y$. In addition, one can find
that they have to commute with the elements of the $R_G$, subgroup
of G. Thus, the four-dimensional gauge group H is the centralizer of
R in G, $H=C_G(R_G)$. The $A_{\alpha}(x,y)$ components of $A_M(x,y)$
denoted by $\phi_{\alpha}(x,y)$ from now on, become scalars in four
dimensions and they transform under R as a vector $\upsilon$, i.e.

\begin{gather}
S\supset R \\
adjS=adjR+\upsilon
\end{gather}
Furthermore, $\phi_{\alpha}(x,y)$ act as an interwining operator
connecting induced representations of $R$ acting on $G$ and $S/R$.
This implies, according to Schur's lemma, that the transformation
properties of the fields $\phi_{\alpha}(x,y)$ under $H$ can be
found, if we express the adjoint representation of $G$ in terms of
$R_G\times H$:

\begin{gather}
G\supset R_G\times H \\
adjG=(adjR,1)+(1,adjH)+\sum(r_i,h_i)
\end{gather}
Then, if $\upsilon =\sum s_i$, where each $s_i$ is an irreducible
representation of $R$, there survives a Higgs multiplet transforming
under the representation $h_i$ of $H$. All other scalar fields
vanish.

The analysis of the constraints imposed on spinors \cite{csdr11,
Manton, Chapline-Slansky,Wetterich-Palla} is analogous to the scalar
cases and implies that the spinor fields act as interwining
operators connecting induced representations of $R$ in $SO(d)$ and
in $G$. In order to specify the representation of H under which the
four-dimensional fermions transform, we have to decompose the
representation $F$ of the initial gauge group in which the fermions
are assigned in higher dimensions under $R_G\times H$, i.e.

\begin{align}
\begin{split}
F=\sum (t_i,h_i)
\end{split}
\end{align}
and the spinor of $SO(d)$ under $R$
\begin{align}
\begin{split}
\sigma_d=\sum \sigma _j
\end{split}
\end{align}
It turns out that for each pair $(r_i,\sigma_i)$, where $r_i$ and
$\sigma_i$ are identical irreducible representations of $R$, there
is an $h_i$ multiplet of spinor fields in the four-dimensional
theory. Regarding the existence of chiral fermions in the effective
theory, we notice that if we start with Dirac fermions in higher
dimensions it is impossible to obtain chiral fermions in four
dimensions. Further requirements must be imposed in order to achieve
chiral fermions in the resulting theory. Imposing the Weyl condition
in $D$ dimensions, we obtain two sets of Weyl fermions with the same
quantum numbers under $H$. This is already a chiral theory, but
still one can go further and try to impose Majorana condition in
order to eliminate the doubling of the fermionic spectrum. Majorana
and Weyl conditions are compatible in $D=4n+2$, which is the case of
our interest.

An important requirement is that the resulting four-dimensional
theories should be anomaly free. Starting with an anomaly free
theory in higher dimensions, Witten \cite {witten} has given the
condition to be fulfilled in order to obtain anomaly free
four-dimensional theories. The condition restricts the allowed
embeddings of $R$ into $G$ by relating them with the embedding of
$R$ into $SO(6)$, the tangent space of the six-dimensional cosets we
consider \cite {csdr11,Piltch}. According to ref. \cite {Piltch} the
anomaly cancellation condition is automatically satisfied for the
choice of embedding
\begin{align}
\begin{split}
E_8\supset SO(6)\supset R,
\end{split}
\end{align}
which we adopt here.

\subsection{Dimensional Reduction of $E_8$ over $SU(3)/U(1)\times
U(1)$}

Let us next present a few results concerning the dimensional
reduction of the $\mathcal{N}=1, E_8$ SYM over $SU(3)/U(1)\times
U(1)$ \cite {Lust-Zoupanos}. To determine the four-dimensional gauge
group, the embedding of $R=U(1)\times U(1)$ in $E_8$ is suggested by
the decomposition

\begin{align}
\begin{split}
E_8\supset E_6\times SU(3)\supset E_6\times U(1)_A\times U(1)_B
\end{split}
\end{align}
After the dimensional reduction of $E_8$ under $SU(3)/U(1)\times
U(1)$, according to the rules of the previous section, the surviving
gauge group in four dimensions is
\begin{align}
\begin{split}
H=C_{E_8}(U(1)_A\times U(1)_B)=E_6\times U(1)_A\times U(1)_B
\end{split}
\end{align}
Similarly, the explicit decomposition of the adjoint representation
of $E_8$, $248$ under $U(1)_A\times U(1)_B$ provides us with the
surviving scalars and fermions in four dimensions. Eventually, one
finds that the dimensionally reduced theory in four dimensions is a
$\mathcal{N}=1$, $E_6$ GUT with $U(1)_A, U(1)_B$ as global
symmetries. The potential is determined by a tedious calculation
\cite{manousselis1, manousselis2}.
The D-terms can be constructed and the F-terms are obtained by the
superpotential. The rest of the terms in the potential could be
interpreted as soft scalar masses and trilinear soft terms. Finally,
the gaugino mass was also calculated and receives contribution from
the torsion contrary to the rest soft supersymmetry breaking terms.

\subsection{$SU(3)^3$ due to Wilson flux}

In order to reduce further the gauge symmetry, one has to apply the
Wilson flux breaking mechanism \cite {Kozimirov,Zoup,Hosotani}.
Instead of considering a gauge theory on $M^4\times B_0$ ($B_0$ a
simply connected manifold in our case), one considers a gauge theory
on $M^4\times B$, with $B=B_0/F^{S/R}$ and $F^{S/R}$ a freely acting
discrete symmetry of $B_0$. The discrete symmetries $F^{S/R}$, which
act freely on coset spaces $B_0=S/R$, are the center of $S$, $Z(S)$
and $W=W_S/W_R$, where $W_S$ and $W_R$ are the Weyl groups of $S$
and $R$, respectively. In the case of our interest \vspace*{0.3cm}
\begin{align}
\begin{split}
F^{S/R}=\mathbb{Z}_{3}\subseteq W
\end{split}
\end{align}
The presence of the Wilson lines imposes further constraints on the
fields of the theory. The surviving fields are invariant under the
combined action of the discrete group $\mathbb{Z}_{3}$ on the
geometry and on the gauge indices.

After the $\mathbb{Z}_{3}$ projection, the gauge group $E_6$ breaks
to \vspace*{0.3cm}
\begin{align}
\begin{split}
SU(3)_C\times SU(3)_L\times SU(3)_R
\end{split}
\end{align}
(the first of the $SU(3)$ factors is the Standard Model colour gauge
group). Moreover, one can obtain three fermion generations by
introducing non-trivial monopole charges in the $U(1)$'s in $R$.

In ref \cite{Irges-Zoupanos} it was shown that the scalar potential
leads to the proper hierarchy of spontaneous breaking. Using the
appropriate vev's, a first spontaneous symmetry breaking leads to
the MSSM \cite{Babu-Pakvasa}, while the electroweak breaking
proceeds by a second one \cite{ma19}. It is worth noting that before
the EW symmetry breaking, supersymmetry is broken by both D-terms
and F-terms, in addition to its breaking by the soft terms.

We plan to examine in detail the phenomenological consequences of
the resulting model, taking also into account the massive Kaluza
Klein modes.

\section{Field theory orbifolds and fuzzy spheres}

Let us begin with reminding briefly how the orbifold structure
applies in field theory, and how this structure is related to the
dynamical generation of fuzzy extra dimensions. The reason is that
we seek to end up with chiral fermions in the case of construction
of models in particle physics.

We commence with a $SU(3N)$ $\mathcal{N}=4$ supersymmetric
Yang-Mills (SYM) theory. The orbifold projection of this theory will
be achieved by the action of the (discrete) group $\mathbb{Z}_3$.
The procedure is to embed a discrete symmetry into the R-symmetry of
the original theory, i.e. $SU(4)_R$. Due to this embedding, the
projected theories we may end up with, may have different amount of
remnant supersymmetry \cite{kachru24}. For example, supersymmetry is
completely broken when $\mathbb{Z}_3$ is embedded maximally in
$SU(4)_R$, while if it is embedded in an $SU(3)$ or $SU(2)$ subgroup
of the $SU(4)_R$, it results to $\mathcal{N}=1$ or $\mathcal{N}=2$
supersymmetric theories, respectively. In this contribution, we
concentrate on the $\mathcal{N}=1$ case, which is compatible with
our prime motivation, that is the construction of chiral models.

Projecting the initial theory under the discrete symmetry
$\mathbb{Z}_3$ leads to a $\mathcal{N}=1$ SYM theory, in which the
only fields that remain are the ones that are invariant under the
action of the discrete group, $\mathbb{Z}_3$. For the technicalities
of this procedure see \cite{chatzistavrakidis17}. In the initial
$\mathcal{N}=4$ SYM theory, there are totally four superfields, one
vector and three chiral in $\mathcal{N}=1$ language. The component
fields are the gauge fields $A_\mu, \mu=0,\ldots3$ of the $SU(3N)$
gauge group, three complex scalar fields $\phi^i, i=1,\ldots3$,
which are accommodated in the adjoint of the gauge group and in the
vector of the global symmetry and four Majorana fermions $\psi^p$,
which are assigned in the adjoint of the gauge group and the spinor
of the
global symmetry. 
After the orbifold pojection, we end up with a theory which has
different gauge group and particle spectrum. In short,
$\mathbb{Z}_3$ acts non-trivially on the various fields depending on
their representations under the R-symmetry and the gauge group
\cite{kachru24}. The gauge group breaks down to $H=S(U(N)\times
U(N)\times U(N))$ and the scalar and fermionic fields that survive,
transform under the representations
\begin{equation}
  3\cdot\left((N\bar{N},1)+(\bar{N},1,N)+(1,N,\bar{N})\right)
\end{equation}
of the non-Abelian factor gauge groups, obtaining a spectrum free of
gauge anomalies. It is easily understood that fermions belong to
chiral representations and that there is a threefold replication,
meaning there are three chiral families.

As for the F-term scalar potential of the $N=4$ SYM theory, we
obtain
\begin{equation}
  V_F(\phi)=\frac{1}{4}\text{Tr}\left(\left[\phi^i,\phi^j\right]^\dag\left[\phi^i,\phi^j\right]\right)\,.
\end{equation}
After the projection, the potential $V_F$ remains practically the
same, obviously containing only the terms which describe
interactions of the surviving fields. We also have a contribution to
the total scalar potential from the D-terms, that is
\begin{equation}
  V_D=\frac{1}{2}D^2=\frac{1}{2}D^ID_I\,,
\end{equation}
with the D-terms having the form
\begin{equation}
  D^I=\phi_i^\dag T^I\phi^i\,,
\end{equation}
where $T^I$ are the generators in the representation of the
corresponding chiral multiplets. Obviously, vanishing both F-terms
and D-terms, which means $\left[\phi^i,\phi^j\right]=0$, we obtain
the minimum of the full scalar potential, i.e. all scalar fields
vanish in the vacuum and therefore no spontaneous supersymmetry
breaking takes place. However, interesting vacua are achieved by
inserting soft supersymmetry breaking terms in the theory.
Specifically, the scalar part of the soft supersymmetric breaking
sector is
\begin{equation}
  V_{SSB}=\frac{1}{2}\sum_im_i^2\phi^{i\dag}\phi^i+\frac{1}{2}\sum_{i,j,k}h_{ijk}\phi^i\phi^j\phi^k+h.c.\,,
\end{equation}
which obeys the orbifold symmetry. Therefore, the expression for the
full scalar potential of the theory becomes now
\begin{equation}
  V=V_F+V_D+V_{SSB}\,,
\end{equation}
which can be equivalently written in the form
\begin{equation}
  V=\frac{1}{4}(F^{ij})^\dag F^{ij}+V_D\,,
\end{equation}
for suitable parameters, having also defined
\begin{equation}
  F^{ij}=\left[\phi^i,\phi^j\right]-i\varepsilon^{ijk}(\phi^k)^\dag\,.
\end{equation}

Due to the fact that the first term is always positive, in order to
obtain the global minimum of the potential, the following equations
must hold
\begin{align}
  &\left[\phi^i,\phi^j\right]=i\varepsilon_{ijk}(\phi^k)^\dag\,,\label{mia}\\
  &\left[(\phi^i)^\dag,(\phi^j)^\dag\right]=i\varepsilon_{ijk}\phi^k\,,\\
  &\phi^i(\phi^i)^\dag=R^2\,,\label{tria}
\end{align}
where $(\phi^i)^\dag$ is the hermitian conjugate of the $\phi^i$ and
$[R^2,\phi^i]=0$. The above relations are related to the fuzzy
sphere. This can be easily understood if we consider the twisted
fields $\tilde{\phi}_i$, which are defined by
\begin{equation}
  \phi^i=\Omega\tilde{\phi}^i\,,
\end{equation}
for $\Omega\neq1$, satisfying the relations
\begin{align}
\Omega^3=1\,,\,
[\Omega,\phi^i]&=0\,,\,\Omega^\dag=\Omega^{-1}\,,\nonumber\\
(\tilde{\phi}^i)^\dag=\tilde{\phi}^i&\Leftrightarrow
(\phi^i)^\dag=\Omega\phi^i.
\end{align}
Therefore, \eqref{mia} converts to the relation of the ordinary
sphere
\begin{equation}
  \left[\tilde{\phi}^i,\tilde{\phi}^j\right]=i\varepsilon_{ijk}\tilde{\phi}^k\,,
\end{equation}
which is generated by $\tilde{\phi}^i$ and \eqref{tria} becomes
\begin{equation}
  \tilde{\phi}^i\tilde{\phi}^i=R^2\,.
\end{equation}
Expressions of $\phi^i$ which satisfy \eqref{mia} have the following
form
\begin{equation}
  \phi^i=\Omega(\mathbf{1}_3\otimes\lambda^i_{(N)})\,,\label{tessera}
\end{equation}
where $\lambda^i_{(N)}$ are the generators of the $SU(2)$ group in
the $N-$dimensional representation. The matrix $\Omega$ is
\begin{equation}
  \Omega=U\otimes\mathbf{1}_N\,,\qquad U=\left(
                                           \begin{array}{ccc}
                                             0 & 1 & 0 \\
                                             0 & 0 & 1 \\
                                             1 & 0 & 0 \\
                                           \end{array}
                                         \right)\,,\qquad
                                         U^3=\mathbf{1}\,.
\end{equation}
The true meaning of the above configuration is revealed by
diagonalizing the matrix $\Omega$, that is
\begin{equation}
  \tilde{\Omega}_3:=U^{-1}\Omega U=\text{diag}(1,\omega,\omega^2)\,.
\end{equation}
Therefore, \eqref{tessera} now becomes
\begin{equation}
\phi^i=\left(
         \begin{array}{ccc}
           \lambda_{(N)}^i & 0 & 0 \\
           0 & \omega\lambda_{(N)}^i & 0 \\
           0 & 0 & \omega^2\lambda_{(N)}^i \\
         \end{array}
       \right)\,.
\end{equation}
This form of $\phi^i$ indicates that there are actually three
identical fuzzy spheres, which are embedded with relative angles
$2\pi/3$.

The solution, \eqref{tessera}, breaks completely the gauge symmetry
$SU(N)^3$ (it could be considered as the Higgs mechanism of the SYM
theory), yet there exists a class of solutions which do not break
the gauge symmetry completely, namely
\begin{equation}
  \phi^i=\Omega\left(\mathbf{1}\otimes(\lambda^i_{(N-n)}\oplus\mathbf{0}_n)\right)\,,
\end{equation}
where $\mathbf{0}_n$ is the $n\times n$ matrix with zero entries. In
this case, the gauge symmetry breaks from $SU(N)^3$ to $SU(n)^3$
with the vacuum being interpreted as $\mathbb{R}\times K_F$, with an
internal fuzzy geometry, $K_F$, of a set of twisted fuzzy spheres
(in $\phi^i$ coordinates). It is possible that this kind of vacua
leads to a low-energy theory of high phenomenological interest, see
discussion in \cite{chatzistavrakidis17}.

Summing up, we should emphasize the general picture of the
theoretical model. At very high-scale regime, we have an unbroken
renormalizable gauge theory. After the spontaneous symmetry
breaking, the resulting gauge theory is an $SU(3)^3$ GUT,
accompanied by an unsurprising finite tower of massive Kaluza-Klein
modes. Finally, the trinification $SU(3)^3$ GUT breaks down to MSSM
in the low scale regime.
\section*{Acknowledgement}
This research is implemented under the Research Funding Program
ARISTEIA, Higher Order Calculations and Tools for High Energy
Colliders, HOCTools and the ARISTEIA II, Investigation of certain
higher derivative term field theories and gravity models
(co-financed by the European Union (European Social Fund ESF) and
Greek national funds through the Operational Program Education and
Lifelong Learning of the National Strategic Reference Framework
(NSRF)), as well as the European Unions ITN programme HIGGSTOOLS.

\noindent One of us, G.Z, would like to thank the organizers for the
warm hospitality.
\end{multicols}


\begin{thebibliography}{99}
\bibitem{gsw} M. B. Green, J. H. Schwarz and E. Witten, Cambridge, Uk: Univ.
Pr. ( 1987) 469 P. ( Cambridge Monographs On Mathematical Physics);
D. Lust and S. Theisen, Lect. Notes Phys. 346 (1989) 1; J.
Polchinski, Cambridge, UK: Univ. Pr. (1998) 531 p; K. Becker, M.
Becker and J. H. Schwarz, Cambridge, UK: Cambridge Univ. Pr. (2007)
739 p; E. Kiritsis, Princeton University Press, 2007.
\bibitem{connes}A. Connes, Academic Press 1994;
J. Madore, Lond.Math. Soc. Lect. Note Ser 257, 1 (2000).
\bibitem{protianafora} N. Seiberg and E. Witten,String theory and non commutative geometry, JHEP 9909, 032 (1999) [arXiv:hep-th/9908142].
\bibitem{LopesCardoso:2002hd}
G.~Lopes Cardoso, G.~Curio, G.~Dall'Agata, D.~Lust, P.~Manousselis
and
 G.~Zoupanos,
 Nucl.\ Phys.\ B { 652} (2003) 5
 [hep-th/0211118].

 \bibitem{Becker:2003yv}
 K.~Becker, M.~Becker, K.~Dasgupta and P.~S.~Green,
 JHEP {0304} (2003) 007
 [hep-th/0301161].

 \bibitem{Becker:2003sh}
 K.~Becker, M.~Becker, P.~S.~Green, K.~Dasgupta and E.~Sharpe,
 Nucl.\ Phys.\ B { 678} (2004) 19
 [hep-th/0310058].

 \bibitem{Gurrieri:2004dt}
 S.~Gurrieri, A.~Lukas and A.~Micu,
 Phys.\ Rev.\ D { 70} (2004) 126009
 [hep-th/0408121].
 \bibitem{Benmachiche:2008ma}
 I.~Benmachiche, J.~Louis and D.~Martinez-Pedrera,
 Class.\ Quant.\ Grav.\ { 25} (2008) 135006
 [arXiv:0802.0410 [hep-th]].
 \bibitem{Micu:2004tz}
 A.~Micu,
 Phys.\ Rev.\ D { 70} (2004) 126002
 [hep-th/0409008].

 \bibitem{Frey:2005zz}
 A.~R.~Frey and M.~Lippert,
 Phys.\ Rev.\ D { 72} (2005) 126001
 [hep-th/0507202].
 \bibitem{Manousselis:2005xa}
 P.~Manousselis, N.~Prezas and G.~Zoupanos,
 Nucl.\ Phys.\ B { 739} (2006) 85
 [hep-th/0511122].

 \bibitem{Chatzistavrakidis:2008ii}
 A.~Chatzistavrakidis, P.~Manousselis and G.~Zoupanos,
 Fortsch.\ Phys.\ {57} (2009) 527
 [arXiv:0811.2182 [hep-th]].
 \bibitem{Chatzistavrakidis:2009mh}
 A.~Chatzistavrakidis and G.~Zoupanos,
 JHEP { 0909} (2009) 077
 [arXiv:0905.2398 [hep-th]].

 \bibitem{Dolan:2009nz}
 B.~P.~Dolan and R.~J.~Szabo,
 JHEP { 0908} (2009) 038
 [arXiv:0905.4899 [hep-th]].

 \bibitem{Lechtenfeld:2010dr}
 O.~Lechtenfeld, C.~Nolle and A.~D.~Popov,
 JHEP { 1009} (2010) 074
 [arXiv:1007.0236 [hep-th]].
 \bibitem{Popov:2010rf}
 A.~D.~Popov and R.~J.~Szabo,
 JHEP { 1202} (2012) 033
 [arXiv:1009.3208 [hep-th]].
 \bibitem{Klaput:2011mz}
 M.~Klaput, A.~Lukas and C.~Matti,
 JHEP { 1109} (2011) 100
 [arXiv:1107.3573 [hep-th]].

 \bibitem{Chatzistavrakidis:2012qb}
 A.~Chatzistavrakidis, O.~Lechtenfeld and A.~D.~Popov,
 JHEP { 1204} (2012) 114
 [arXiv:1202.1278 [hep-th]].
 \bibitem{Gray:2012md}
 J.~Gray, M.~Larfors and D.~Lust,
 JHEP { 1208} (2012) 099
 [arXiv:1205.6208 [hep-th]].

\bibitem{Klaput:2012vv}
 M.~Klaput, A.~Lukas, C.~Matti and E.~E.~Svanes,
JHEP { 1301} (2013) 015 [arXiv:1210.5933 [hep-th]].


\bibitem{csdr10} P. Forgacs and N. S. Manton, Commun. Math. Phys. 72 (1980)
15; E.~Witten, Phys. Rev. Lett.\ 38, 121 (1977).
\bibitem{csdr11} D. Kapetanakis and G. Zoupanos, Phys. Rept. 219 (1992) 1.
\bibitem{csdr12} Y. .A. Kubyshin, I. P. Volobuev, J. M. Mourao and G. Rudolph, Lect. Notes Phys. 349 (1990) 1.
\bibitem{scherk-schwarz}
J.Scherk, J.H.Schwarz, Phys.Lett. B82 (1979) 60
\bibitem{CY} P. Candelas, G. T. Horowitz, A. Strominger, E. Witten, Nucl.
Phys.B 258, (1985) 46
\bibitem{grana} M. Grana, Phys. Rept. 423, 91 (2006), arXiv: 0509003[hep-th]; M.R.
  Douglas, S. Kachru, Rev. Mod.
  Phys. 79 (2007) 733 [hep-th/0610102]; R. Blumenhagen, B. Kors, D. Lόst, S.
  Stieberger, Phys. Rept. 445
  (2007) 1 [hep-th/0610327]; F. Denef, arXiv:0803.1194 [hep-th]
\bibitem{Butruille}
  Butruille J. -B.,
[arXiv:math.DG/0612655].
\bibitem{Irges-Zoupanos}
  N.Irges and G.Zoupanos,
Phys.\ Lett.\ B698, (2011) 146 [arXiv:hep-ph/1102.2220]; N.Irges,
G.Orfanidis, G.Zoupanos, arXiv:1205.0753 [hep-ph].
\bibitem{filk6} T. Filk, Phys. Lett. B 376 (1996) 53; J. C. Varilly and J. M.
Gracia-Bondia, Int. J. Mod. Phys. A 14 (1999) 1305 [hep-th/9804001];
M. Chaichian, A. Demichev and P. Presnajder, Nucl. Phys. B 567
(2000) 360 [hep-th/9812180]; S. Minwalla, M. Van Raamsdonk and N.
Seiberg, JHEP 0002 (2000) 020 [hep-th/9912072].
\bibitem{groose7} H.
Grosse and R. Wulkenhaar, Lett. Math. Phys. 71 (2005) 13
[hep-th/0403232].
\bibitem{groose8} H. Grosse and H. Steinacker, Adv. Theor. Math. Phys. 12
(2008) 605 [hep-th/0607235];H. Grosse and H. Steinacker, Nucl. Phys.
B 707 (2005) 145 [hep-th/0407089].
\bibitem{aschieri9}
P. Aschieri, J. Madore, P. Manousselis and G. Zoupanos, JHEP 0404
(2004) 034 [hep-th/0310072]; P. Aschieri, J. Madore, P. Manousselis
and G. Zoupanos, Fortsch. Phys. 52 (2004) 718 [hep-th/0401200].
\bibitem{schupp} B.Jurco, P.Schupp, J.Wess, Nucl.Phys. B604 (2001)
148-180, [arXiv:hep-th/0102129]
\bibitem{aschieri13}P. Aschieri, T. Grammatikopoulos, H. Steinacker and G. Zoupanos, JHEP 0609 (2006) 026
[hep-th/0606021].
\bibitem{steinacker14}H. Steinacker and G. Zoupanos, JHEP 0709 (2007) 017 [arXiv:0706.0398 [hep-th]].
\bibitem{chatzistavrakidis15}A. Chatzistavrakidis, H. Steinacker and G. Zoupanos, Fortsch. Phys. 58 (2010) 537 [arXiv:0909.5559
[hep-th]].
\bibitem{maalampi16}J. Maalampi and M. Roos, Phys. Rept. 186 (1990) 53.
\bibitem{chatzistavrakidis17}A. Chatzistavrakidis, H. Steinacker and G. Zoupanos,
JHEP 1005 (2010) 100 [arXiv:1002.2606 [hep-th]].
\bibitem{inspire2} S. Heinemeyer,M.Mondragon and G. Zoupanos,
Int.J.Mod.Phys. A29 (2014) 18, [hep-ph/1430032].
\bibitem{inspire4} M. Mondragon,N.Tracas and G. Zoupanos [arXiv:1403.7384
[hep-ph]].
\bibitem{heinemeyer18}S. Heinemeyer,M. Mondragon and G. Zoupanos, SIGMA 6 (2010) 049 [arXiv:1001.0428 [hep-ph]].
\bibitem{ma19}E. Ma, M. Mondragon and G. Zoupanos, JHEP 0412 (2004) 026 [hep-ph/0407236];
S. Heinemeyer, E. Ma, M. Mondragon and G. Zoupanos, AIP Conf. Proc.
1200 (2010) 568 [arXiv:0910.0501 [hep-ph]].
\bibitem{Manton}
  N.S.~Manton.
Nucl.\ Phys.\  B193, 502(1981).
\bibitem{Chapline-Slansky}
  G.~Chapline and R.~Slansky.
Nucl.\ Phys.\  B204, 461(1982).
\bibitem{Wetterich-Palla}
  C.~Wetterich, Nucl.\ Phys.\  B222, 20(1985); L.~Palla, Z. Phys. \ C24, 345(1983); K.~Pilch and A.N.~Schellekens, J.~Math. Phys. 25, 3455(1984);
  P.~Forgacs, Z.~Horvath and L.~Palla, Z. Phys C30, 261(1986); K.J.~Barnes, P.~Forgacs, M.~Surridge and G.~Zoupanos, Z. Phys. C33, 427(1987).
\bibitem{witten}
E. Witten, Phys. Lett. B144, 351(1984).
\bibitem{Piltch}
K. Pilch and A. N. Schellekens, Nucl. Phys. B259, 673(1985); D.
Luest, Nucl. Phys. B276, 220(1985); D. Kapetanakis and G. Zoupanos,
Phys. Lett. B249, 66(1990).
\bibitem{Lust-Zoupanos}
D. L$\ddot {u}$st and G. Zoupanos, Phys. Lett. B165 (1985) 309; D.
Kapetanakis and G. Zoupanos, Z. Phys. C56 (1992); G. Douzas, T.
Grammatikopoulos and G. Zoupanos, Eur. Phys. J. C59 (2009) 917
\bibitem{manousselis1}
P.Manousselis and G. Zoupanos, JHEP 0203 (2002) 002.
\bibitem{manousselis2}
P. Manousselis and G. Zoupanos, JHEP 0411 (2004) 025.
\bibitem{Kozimirov}
N. Kozimirov, V. A. Kuzmin and I. I. Tkachev, Sov. J. Nucl. Phys.
49, 164(1989); D. Kapetanakis and G. Zoupanos, Phys. Lett. B232,
104(1989).
\bibitem{Zoup}
G. Zoupanos, Phys. Lett. B201, 301(1988).
\bibitem{Hosotani}
Y. Hosotani, Phys. Lett. B126, 309(1983); B129, 193(1983); E.
Witten, Nucl. Phys B258, 75(1985); J. D. Breit, B. A. Ovrut and G.
C. Segre, Phys. Lett. B158, 33(1985); B. Greene, K. Kirklin and P.
J. Miron, Nucl. Phys. B274, 574(1986); B. Greene, K. Kirklin, P. J.
Miron and G. G. Ross, Nucl. Phys. B278, 667(1986).
\bibitem{Babu-Pakvasa}
K.S. Babu, X.-G. He and S. Pakvasa, Phys. Rev. D33 (1986) 763; G.K.
Leontaris and J. Rizos, Phys. Lett. B632 (2006) 710; J. Sayre, S.
Wiesenfeld and S. Willenbrock, Phys. Rev. D73 (2006) 035013.
\bibitem{kachru24}S. Kachru and E. Silverstein, Phys. Rev. Lett. 80 (1998) 4855 [hep-th/9802183].
\end{thebibliography}
\end{document}